# Ideal type-II Weyl phase and topological transition in phononic crystals

Xueqin Huang[1], Jiuyang Lu[1], Feng Li[1], Weiyin Deng[1,†], Zhengyou Liu[2,3,†]

[1]School of Physics and Optoelectronics, South China University of Technology, Guangzhou, Guangdong 510640, China
[2]Key Laboratory of Artificial Micro- and Nanostructures of Ministry of Education and School of Physics and Technology, Wuhan University, Wuhan 430072, China
[3]Institute for Advanced Studies, Wuhan University, Wuhan 430072, China

†Corresponding author. Email: dengwy@scut.edu.cn; zyliu@whu.edu.cn

**Abstract**: Ideal Weyl points, which are related by symmetry and thus reside at the same frequency, could promote the deep development and utilization of the Weyl physics. Although the ideal type-I Weyl points have been achieved in photonic crystals, the ideal type-II Weyl points with tilted cone-like band dispersions, are still beyond discovery. Here we realize ideal type-II Weyl points of minimal number in three-dimensional layer-stacked phononic crystals, and demonstrate topological phase transition from Weyl semimetal to valley insulators of two distinct types. The Fermi-arc surface states exist on the interface of the Weyl and valley phases, while the Fermi-circle ones occur on that of the two distinct valley phases. We show the interesting wave partition of Fermi-circle surface states on the interfaces formed by distinct valley phases.

Weyl semimetals, hosting twofold linear crossing points in three-dimensional (3D) momentum space, i.e. Weyl points (WPs), have stimulated intensive and innovative research in the field of topological physics [1-8]. The WP behaves as a source or sink of the Berry curvature flux, and carries topological charge with a value of $+1$ or $-1$ [3]. The existence of WPs needs to break the spatial inversion or time-reversal symmetries. Generally, there exists at least two WPs in the system with time-reversal symmetry broken, while the minimal number is four in the presence of time-reversal symmetry [9]. Because of their topological nature, WPs coming in pairs give rise to a variety of intriguing properties, such as nonclosed Fermi-arc surface dispersions [1-3] and chiral anomaly [10-15]. In particular, WPs described by all the three Pauli matrices are stable against small perturbations [16]. Therefore, the topological phase transition in Weyl semimetal is much less explored.

There are two types of WPs: the type-I WP has a point-like Fermi surface, while the type-II WP hosts a strongly tilted cone dispersion and possess a conical Fermi surface [17-20]. This feature enables the type-II Weyl semimetal to exhibit numerous unique properties, including anisotropic chiral anomaly [21, 22] and antichiral Landau level [17, 23]. The WPs, of both the type-I and type-II, have been achieved in condensed matter systems [24-31] and artificial periodical structures, such as photonic crystals [32-36] and phononic crystals (PCs) [37-43]. However, these WPs are not at the same energy, making a serious restriction for realizing and utilizing the unique properties induced by the WPs [9, 44-46]. To overcome this problem, recently, the concept of ideal WPs, which are related by symmetry and thus reside at the same energy, is proposed [9, 44-46]. But all of the WPs referred to, belong to the first type or type-I, and therefore the ideal WPs of the second type or type-II are still beyond discovery.

In this work, we design a 3D PC, which hosts four ideal type-II WPs protected by the time-reversal and mirror symmetries. Phase transitions from the Weyl semimetal to valley insulators of two distinct phases are studied. The Fermi-arc surface states are shown to exist on the interface of the Weyl and valley phases. Specifically, the Fermi-circle surface states are found on the interface of the two distinct valley phases. A unique wave partition of the Fermi-circle surface states is demonstrated. All the simulations are performed by the COMSOL Multiphysics solver package. As schematic in Fig. 1(a), the PC is a layer-by-layer stacking of hexagonal array of triangular scatterers, together with the supporting plate. The plate is drilled with cutting-through holes of two different sizes, as shown in Figs. 1(a)-1(c), by which the neighboring layer couplings. The side and top views of one unit cell are displayed in Figs. 1(b) and 1(c). The rotational angle of the triangular rod with respect to the $x$ axis is $\theta$. The height and side length of the triangular rod are $h_1$ and s. The height and the radii of the coupling holes is $h_2$, $r_1$ and $r_2$. The lattice constants in $x$-$y$ plane and along the $z$ direction are $a$ and $h = 2h_1 + h_2$. The parameters are chosen as $a = \sqrt{3}$cm, $s = 0.4\sqrt{3}a$, $h_1 = 0.25a$, $h_2 = 0.1a$, $r_1 = 0.1a$, and $r_2 = 0.3r_1$. Air is filled inside the structured channels with hard boundaries marked by cream color. The mass density of air is chosen as $\rho = 1.3$ kg/m$^3$ and the speed of sound is $v = 343$ m/s. For $\theta = 35°$, the bulk band structure of the lowest two bands at $(k_x, k_y) = (\frac{2\pi}{\sqrt{3}a}, -\frac{2\pi}{3a})$ along the $k_z$ direction is calculated in Fig. 1(e). It is found that two linear crossing points with the same frequency are located at $k_z$ with opposite values. At each one, the two bands possess group velocities in the same direction. For instance, two bands have positive group velocities for $k_z > 0$. To study these crossing points, the bulk band structure in the $k_x$-$k_y$ plane with $k_z = 0.57\,\pi/h$ is plotted in Fig. 1(f). Linear dispersion appears at the $\overline{K}$ point, but the group velocities are in the opposite directions. Therefore, these linear crossing points are type-II WPs with positions at $(k_x, k_y, k_z) = (\frac{2\pi}{\sqrt{3}a}, \pm\frac{2\pi}{3a}, \pm 0.57\frac{\pi}{h})$.

Figure 1(d) shows the schematic of the type-II WP distributions in the first Brillouin zone (BZ) with green and purple spheres denoting opposite charges. There are a minimal number of four type-II WPs in the first BZ. These type-II WPs are ideal because of all of them being at the same frequency, which are guaranteed by the time-reversal symmetry and mirror symmetry. The time-reversal symmetry protects the WPs situated at the $\overline{K}$ and $\overline{K'}$ points having opposite charges, and the mirror symmetry of the $z = 0$ plane protects the WPs situated at opposite $k_z$ also having opposite charges. In addition, owing to the $D_{3h}$ symmetry, the WPs are usually located along the KH and K'H' lines, making the WPs more easily searching.

To study the evolution of these type-II WPs in our PCs, we calculate the bulk band structures at $(k_x, k_y) = (\frac{2\pi}{\sqrt{3}a}, -\frac{2\pi}{3a})$ as functions of $k_z$ and $\theta$, as plotted in Fig. 2(a). When $\theta$ is in the range of $29.1° < \theta < 39°$, type-II WPs emerge as marked by the black dashed line. When $\theta < 29.1°$, the WPs eliminate at $k_z = 0$ and turn to open a bulk gap, while $\theta > 39°$ the WPs eliminate at $k_z = \pm\pi/h$ and open a bulk gap. The gaps for $\theta < 29.1°$ and $\theta > 39°$ should have different topological phases because of the separation of them being the WPs. So there exists three topologically distinct phases I, II and III, colored by gray, blue and red in Fig. 2(b). Because of the $D_{3h}$ symmetry, the band structure as a function of $\theta$ satisfies the periodicity of $120°$. As the system has the mirror symmetry of the $y = 0$ plane at $\theta = 0°$, the topological phase for $\theta < 0°$ keeps the same as that for its positive counterpart, thus only the topological phase for $0° < \theta < 60°$ is shown in Fig. 2. The topological properties of this system can be described by the $k_z$-dependent valley Chern numbers $C_v^K(k_z)$ and $C_v^{K'}(k_z)$, defined in the reduced two-dimensional BZ by considering $k_z$ as a parameter. Time-reversal symmetry guarantees $C_v^K(k_z) = -C_v^{K'}(k_z)$. Phases I and III are the valley insulators, in which the valley Chern numbers of the $\overline{K}$ point are $C_v^K = \frac{1}{2}$ and $-\frac{1}{2}$, respectively. Phase II is the Weyl semimetal, where the WP at the $\overline{K}$ point hosts the topological charge $C_{\overline{K}} = C_v^K(k_z = \overline{K} + 0^+) - C_v^K(k_z = \overline{K} - 0^+) = 1$ (purple dashed line in Fig. 2(b)), while the topological charge of WP at opposite $k_z$ is $-1$ (green dashed line).

Weyl semimetal was predicted to give rise to the Fermi-arc surface states. The Fermi-arc surface states exist at the interface between the phase II and other two phases, because of the nontrivial topology described by $C_v^K(k_z)$. Considering a zigzag interface formed by the PCs with $\theta = 20°$ (phase I) and $\theta = 35°$ (phase II) along the $y$ direction, the projected band structure for $k_z = 0$ is calculated in Fig. 3(a). The projected bulk bands are denoted by the black lines, and the dispersions of surface waves are labelled by the red lines. This is consistent with the phase diagram shown in Fig. 3(d), in which these two PCs possess different valley Chern numbers at $k_z = 0$ and the difference between them is $\pm 1$. However, for $k_z = \pi/h$, these two PCs have the same topology resulting in no surface states, as shown in Fig. 3(b). The Fermi arc, i.e., equi-frequency contour, at the frequency $f = 10$ kHz is plotted in Fig. 3(c). The green and purple spheres are the projections of the WPs with opposite topological charges. Although merging with the projected bulk band (the black lines) due to the tilted dispersion of type-II WPs, the Fermi arc (red lines) connected the two WPs with opposite charges are clearly to be revealed because of the ideal property. Interestingly, for the interface between the PCs with $\theta = 35°$ (phase II) and $\theta = 50°$ (phase III), the surface dispersions exist at $k_z = \pi/h$, rather than $k_z = 0$, as shown in Figs. 3(e) and 3(f), for the same topological reason (Fig. 3(h)). The Fermi arc connects the two WPs by a new way, as shown in Fig. 3(g). Therefore, the configuration of Fermi arc can be manipulated by changing the interface.

Finally, we study the surface states along the interface between the two valley insulators. Figures 4(a)-4(b) show the equi-frequency contours at $f = 10.7$ kHz for two different interfaces formed by the PCs with $\theta = 20°$ (phase I) and $\theta = 50°$ (phase III). For the two distinct valley insulators, where the valley Chern numbers are opposite for all $k_z$, the surface wave dispersions form the Fermi circle configuration, which are mainly determined by the layer couplings. Compared with the Fermi-arc surface states merged with the bulk states, the Fermi-circle surface states can be existed in the bulk gap and have the advantage in the wave transport. A point source can excite them without being afraid of the projected bulk states. As a concrete example, we display an interesting wave partition of the surface states at the intersection channels. Figure 4(c) shows schematic of the system constructed by four domains with channels 1 to 4. The simulated field distributions of wave partition are shown in Fig. 4(d). The directions of group velocities in channels 1, 2, 4 are shown in Fig. 4(a), while those in channel 3 are shown in Fig. 4(b). The incident waves from channel 1 only enter into channel 2 and 4, because of the valley-momentum locking. Dramatically, the sound waves mainly propagate to channel 4 with a sharp turn, since the surface states near a sharp turn can be coupled more easily, similar to the partition of the edge states in the two-dimensional topological system [47-49]. The surface waves also split into two parts along the $z$ direction, coming from the specific dispersions.

In conclusion, we realize by design for the first time the ideal type-II WPs, which provides special advantages to explore the intriguing properties of type-II Weyl semimetals, compared with conventional ones. For example, the antichiral effect of the Landau levels may be observed more easily in ideal type-II Weyl semimetals. The annihilation of the WPs gives rise to the valley insulating phases, which offer the unique Fermi-circle surface states for wave manipulation, such as wave partition. As other 3D topological PCs [39-43, 50-52], our structures should be easy to fabricate by common 3D printing technique, allowing the experimental observations of the ideal type-II WPs, the transition to the valley insulators, and the relevant surface states.

**Acknowledgements**

This work is supported by the National Natural Science Foundation of China (Grant Nos. 11804101, 11890701, 11572318, 11604102, 11704128, 11774275, 11974120 and 11974005), the National Key R&D Program of China (Grant No. 2018FYA0305800), Guangdong Innovative and Entrepreneurial Research Team Program (Grant No. 2016ZT06C594), and the Fundamental Research Funds for the Central Universities (Grant Nos. 2018MS93, 2019JQ07, and 2019ZD49).

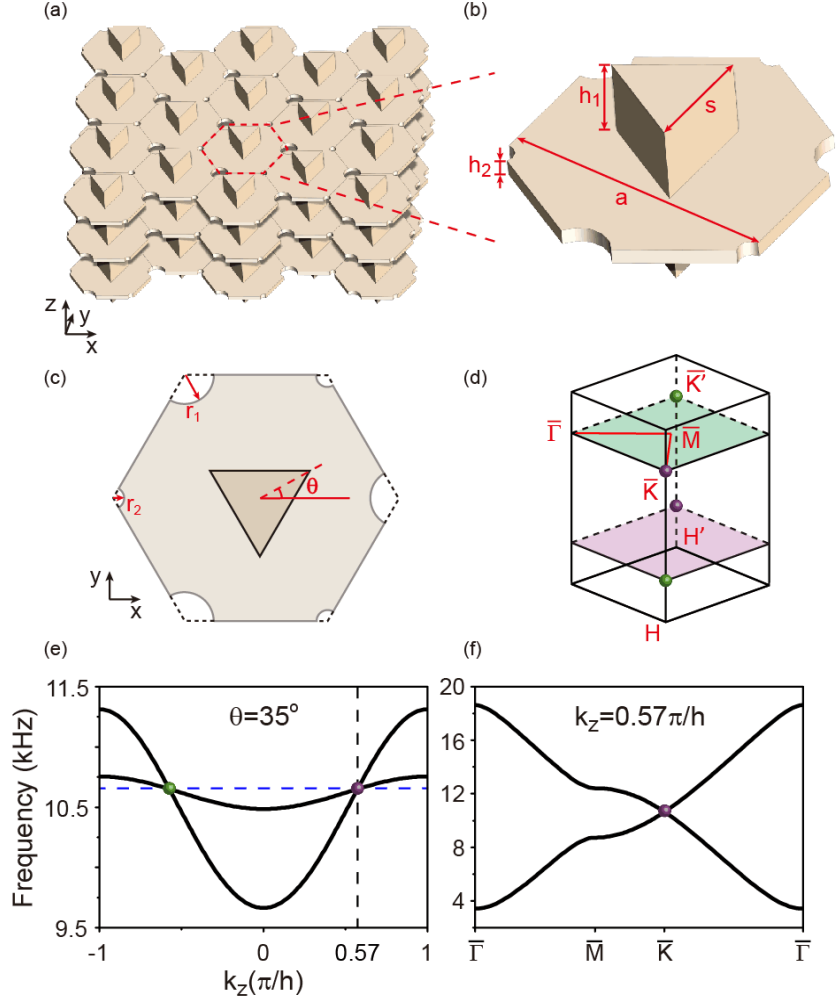

FIG. 1. (a) Schematic of the layer-by-layer stacking phononic crystal. Each layer consists of triangular rods arranged in a hexagonal lattice and couples with each other via two different sized vertical holes. The side (b) and top (c) views of the unit cell. $\theta$ is the rotation angle of the triangular rod with respect to *x* axis. (d) The reduced first Brillouin zone. For $\theta = 35°$, the bulk band structures at $(k_x, k_y) = (\frac{2\pi}{\sqrt{3}a}, -\frac{2\pi}{3a})$ along the $k_z$ direction (e) and in the $k_x$-$k_y$ plane with $k_z = 0.57\frac{\pi}{h}$ (f). There exists a minimal number of four ideal type-II WPs with positions shown in (d). The purple and green circles represent the WPs with charges $+1$ and $-1$, respectively.

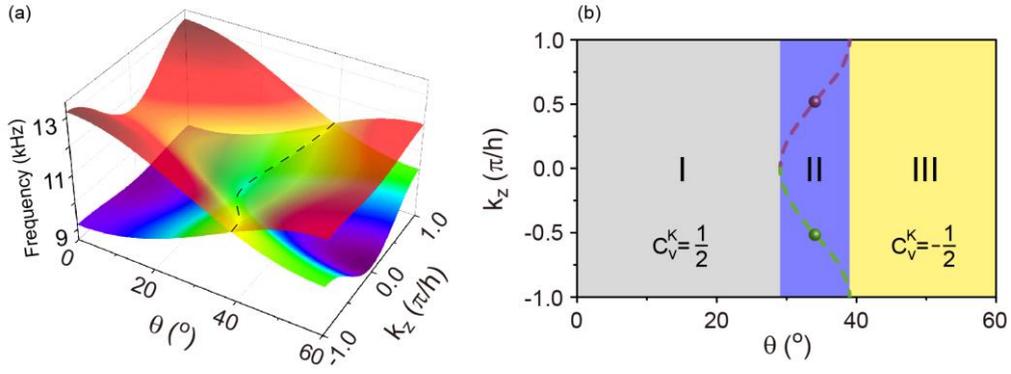

FIG. 2. (a) 3D band structures as functions of $k_z$ and rotation angle $\theta$ at $(k_x, k_y) = (\frac{2\pi}{\sqrt{3}a}, -\frac{2\pi}{3a})$. The dashed line represents the existence of WPs with closing the band gap. (b) The phase diagram determined by the $k_z$-dependent valley Chern numbers for $\theta$. Phases I and III are valley insulators, while phase II is Weyl semimetal. The purple and green dashed lines in phase II represent the positions of WPs with opposite charges.

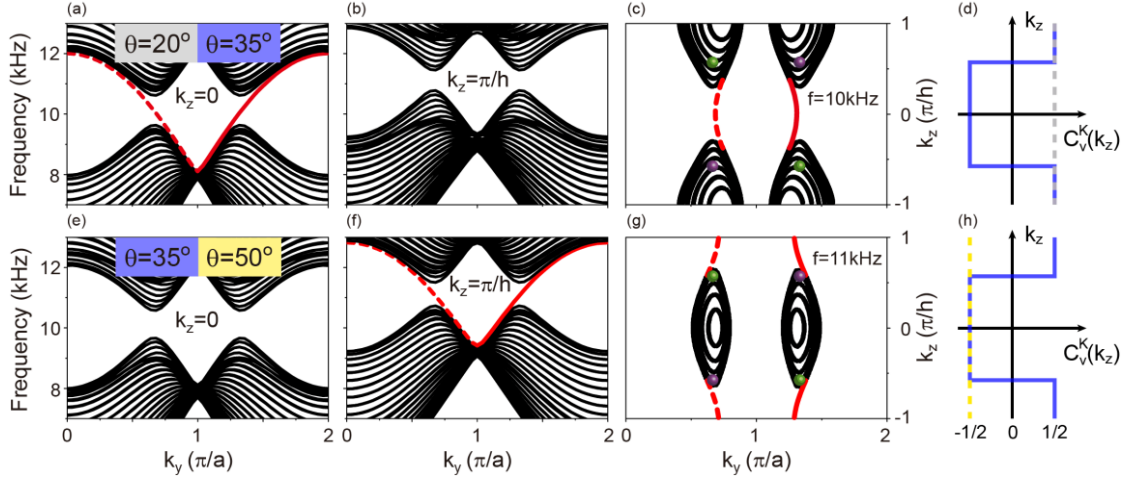

FIG. 3. The surface dispersions with $k_z = 0$ (a) and $k_z = \pi/h$ (b), the equi-frequency contour (c) for the interface formed by the PCs with $\theta = 20°$ (phase I) and $= 35°$ (phase II). The red dashed and solid lines denote the surface states projected from the K′ and K valleys. (d) The valley Chern number $C_v^K(k_z)$ for $\theta = 20°$ (grey dashed line) and $\theta = 35°$ (blue solid line). (e)-(h) The corresponding results for the PCs with $\theta = 35°$ and $\theta = 50°$ (phase III). The Fermi-arc surface dispersions originate from the WPs in Weyl semimetal, but their configurations are related with the valley insulators at the other side of the interface.

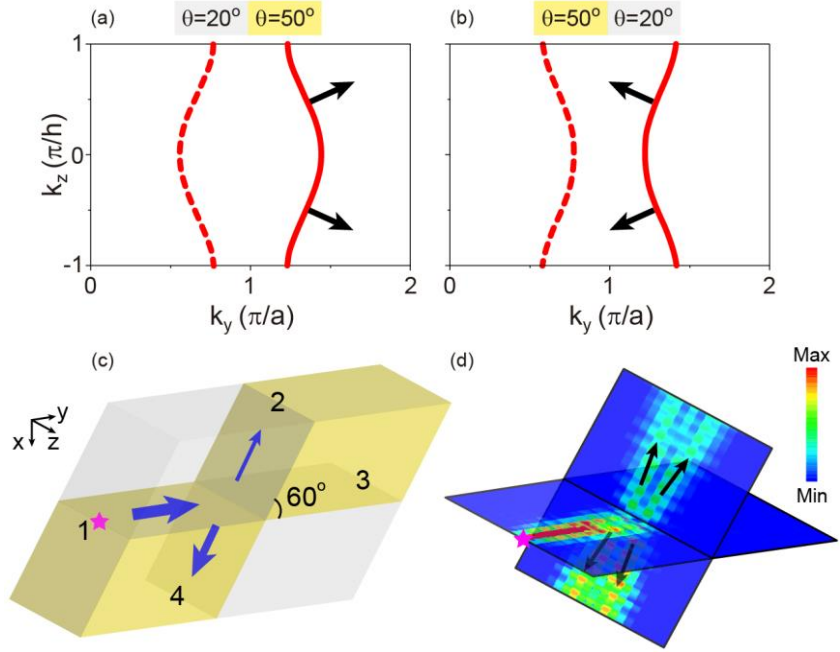

FIG. 4. (a)-(b) The equi-frequency contours for two different interfaces formed by the PCs with $\theta = 20°$ (phase I) and $\theta = 50°$ (phase III). The black arrows denote the direction of the group velocities of surface states projected from K valley. (c) Schematic of the wave partition at an intersection formed by four domains with channels 1 to 4, where the angle between the channels 2 and 3 is $60°$. The incident waves from channel 1 is forbidden to channel 3, and tend to propagate to channel 4 with a sharp turn (shown by the size of blue arrows). (d) The simulated field distributions excited by a point source from channel 1. The frequency is chosen as $f = 10.7\text{kHz}$ in calculation.